\documentclass[traditabstract, letter]{aa}
\usepackage{natbib}
\usepackage{txfonts}
\usepackage{graphicx}
\titlerunning{OGLE-LMC-CEP1812 is a stellar merger}
\authorrunning{Neilson et al.}

\begin{document}

\title{The strange evolution of the Large Magellanic Cloud Cepheid OGLE-LMC-CEP1812}

\author{Hilding R. Neilson\inst{1} \and Robert G.~Izzard\inst{2,3} \and Norbert Langer\inst{3} \and Richard Ignace \inst{4}}
\institute{
Department of Astronomy \& Astrophysics, University of Toronto, 50 St. George Street, Toronto, ON, M5S 3H4, Canada\\
   \email{neilson@astro.utoronto.ca}
   \and
      Institute of Astronomy, University of Cambridge, Madingley Road, Cambridge, CB3 0HA, United Kingdom
      \and
        Argelander-Institut f\"{u}r Astronomie, Universit\"{a}t Bonn, Auf dem H\"{u}gel 71, D-53121 Bonn, Germany 
        \and
   Department of Physics \& Astronomy, East Tennessee State University, Box 70652, Johnson City, TN 37614 USA
  }

\date{}

\abstract{
Classical Cepheids are key probes of both stellar astrophysics and cosmology as standard candles and pulsating variable stars. It is important to understand Cepheids in unprecedented detail in preparation for upcoming GAIA, JWST and extremely-large telescope observations.  Cepheid eclipsing binary stars are ideal tools for achieving this goal, however there are currently only three known systems.  One of those systems, OGLE-LMC-CEP1812, raises new questions about the evolution of classical Cepheids because of an apparent age discrepancy between the Cepheid and its red giant companion.  We show that the Cepheid component is actually the product of a stellar merger of two main sequence stars that has since evolved across the Hertzsprung gap of the HR diagram.  This post-merger product  appears younger than the companion, hence the apparent age discrepancy is resolved. We discuss this idea and consequences for understanding Cepheid evolution. 

}
\maketitle

\section{Introduction}
Classical Cepheids have been crucial for the understanding of stellar astrophysics and cosmology since the discovery of the Cepheid Leavitt law more than a century ago \citep{Leavitt1908}.   They have been used to measure the Hubble constant \citep{Hubble1929, Riess2011, Freedman2012} to a precision of 2\% as well as to constrain the detailed physics of stellar structure and evolution \citep[e.g.,][]{Bono2000}, yet there still exist a number of mysteries regarding these stars.

One such challenge is the detailed calibration of the Cepheid Leavitt law, i.e., the period-luminosity (PL) relation. In the forthcoming era of the \emph{James Webb Space Telescope}, it is expected that we will be able to measure the Hubble constant to less than 1\% precision \citep{Freedman2010}, but this requires measurements of Cepheid fundamental parameters to unprecedented accuracy along with independent measurements of Cepheid distances.  The \emph{Gaia} satellite is currently operating and will measure distances to thousands of Galactic Cepheids \citep{Windmark2010}, that will complement Large Magellanic Cloud (LMC) Cepheids for which distances are known.   

One enduring uncertainty is the decades-old `Cepheid mass discrepancy' \citep{Cox1980}, where Cepheid masses measured using stellar evolution and pulsation calculations differ by about 10 -- 20\% \citep{Keller2008}.  \cite{Bono2006} suggested four possible resolutions to this discrepancy: changes in the radiative opacities, rotation, convective core overshooting in main sequence progenitors and stellar mass loss.  \cite{Neilson2011} showed that pulsation-driven mass loss can partly explain the mass discrepancy, while \cite{Anderson2014} proposed rotation as an alternative solution.  Convective core overshooting has also been shown to resolve the mass discrepancy \citep{Cassisi2011}.  The solution to the mass discrepancy might simply be a combination of all three and understanding which physical  processes are important will constrain both evolution and pulsation models.  However, an ideal method to constrain the Cepheid mass discrepancy is to independently measure Cepheid masses.

 Cepheid distances and masses can be measured in eclipsing binary systems.  \cite{Evans2005} suggested that about 35\% of all Galactic Cepheids are in spectroscopic binary systems, but none are known to be in eclipsing binary systems.  \cite{Evans2013} presented a catalogue of binary companions detected using ultraviolet spectral observations, where the companions are all about 2~$M_\odot$, inferring a  larger binary fraction of 60\%. \cite{Evans2015} report radial velocity measurements which further refine the measured binary fraction of Cepheids, noting that what has been measured is a minimum possible binary fraction.  \cite{Neilson2015} show that a binary fraction of about 60\% is consistent with the observed binary fraction of main sequence B-type stars from \cite{Kouwenhoven}.  
 
Three eclipsing binary Cepheids have  been discovered in the LMC \citep{Soszynski2008}.  These eclipsing binary systems provide unique mass and distance estimates that can be compared to estimates using Cepheid evolution and pulsation models. These help to resolve the Cepheid mass discrepancy and constrain the PL relation in greater detail.

\cite{Piet2010} presented detailed observations of one such eclipsing binary, OGLE-LMC-CEP0227 (CEP0227), with a 310~day orbit and a mass ratio between the components of $q = 1.00 \pm 0.01$.  One binary component is a classical Cepheid and the companion a red giant star.  The authors measured a Cepheid mass $M = 4.14\pm 0.05~M_\odot$ which is consistent with stellar pulsation estimates.  \cite{Cassisi2011} compared the observed fundamental parameters with stellar evolution calculations and found that evolutionary models agree with measured parameters if one assumes moderate convective core overshooting, a result confirmed by other studies \citep{Neilson2012, Prada2012}. \cite{Pilecki2013} analyzed the binary light curve in greater detail to measure the projection factor that is crucial for using the Baade--Wesselink method to measure distances \citep{Baade1929, Storm2011a, Storm2011b, ngeow2012, Neilson2012t}. This specific eclipsing binary system constrains both the Cepheid mass discrepancy and the calibration of the distance scale.

\cite{Gieren2014} presented new observations of another eclipsing binary OGLE-LMC-CEP1718 that appears to be composed of two equal mass Cepheids, both pulsating in the first overtone.   That system has an orbital period of 413-days, but observations are limited and do not independently constrain each star's radius and mass, making conclusions about the their evolution difficult. This system is appears similar to CEP0227.

While analysis of CEP0227 helps resolve problems, the binary system OGLE-LMC-CEP1812 (CEP1812) appears to create problems. \cite{Piet2011} measured the masses and radii of the two stars in a 551~day orbit, but found that the Cepheid appears to be about 100~Myr younger than its red giant companion.  This result raises questions about the evolution of this binary system.  While the binary CEP0227 is consistent with stellar evolution calculations, the binary CEP1812 is not, even though the Cepheid mass is consistent with pulsation calculations. \cite{Piet2011} hypothesized that the Cepheid captured the red giant companion into a binary orbit at some point during its evolution.

In this article, we hypothesize that the Cepheid star in the system CEP1812 evolved from the merger of two main sequence stars. We discuss our stellar evolution code and models in Sect.~2 and present measurements of the age difference between the two stars in the systems using those models in Sect.~3.  In Sect.~4, we compute stellar evolution models with changing mass consistent with our merger scenario and show that our hypothesis resolves the age discrepancy.  In Sect.~5 we discuss the implications of a stellar merger scenario for understanding Cepheid binaries and Cepheid structure and evolution.

\section{Stellar evolution models}
In this work, we computed stellar evolution models using the \cite{Yoon2005} code.  This code has been used to study massive star evolution, supernova progenitors and gamma-ray bursts \citep[e.g.][]{Cantiello2009, Brott2011}.  We have also used this code to explore the evolution of classical Cepheids \citep{Neilson2012a, Neilson2012b, Neilson2014}. 
Models were computed assuming moderate convective core overshooting, consistent with that measured for the LMC binary Cepheid OGLE-LMC-CEP0227 \citep{Cassisi2011, Neilson2012, Prada2012}.  In this case, we write the convective core overshooting efficiency as $0.2H_P$, that is a fraction of the pressure scale height \citep{Brott2011}. Convective core overshooting during main sequence evolution acts to create a more massive helium core, leading to a more luminous giant star. It also acts to prolong main sequence evolution and changes the measured ages of post-main sequence stars.  The models included the \cite{deJager1988} mass-loss prescription for cool stars and the \cite{Kudritzki1989} recipe for hot stars.  We did not assume enhanced mass loss during the Cepheid stage of evolution \citep{Neilson2012a,Neilson2012b}.  The models  assume a composition based on the \cite{Grevesse1998} solar abundances scaled to the standard LMC metallicity $Z = 0.008$  along with an initial helium abundance $Y = 0.256$ \citep{Brott2011}.

Stellar evolution models are constrained by the measured mass and radius of each component of OGLE-LMC-CEP1812.  The Cepheid has mass $M = 3.74\pm 0.06~M_\odot$ and radius $R = 17.4\pm0.9~R_\odot$, while the red giant companion has mass $M_{\rm{RG}} = 2.64\pm0.4~M_\odot$ and radius $R_{\rm{RG}} = 12.1\pm 2.3~R_\odot$. Unfortunately, \cite{Piet2011} did not measure the effective temperature of either star, so we have fewer constraints for CEP1812 than for CEP0227 \cite[e.g.][]{Cassisi2011}. 

\begin{figure}[t]
\begin{center}
\includegraphics[width=0.5\textwidth]{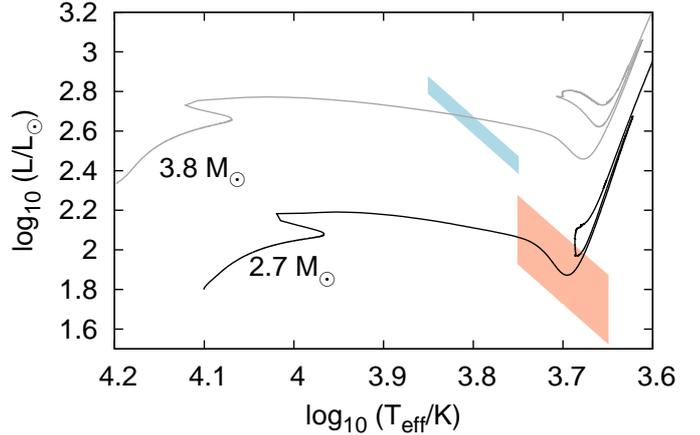}
\caption{Hertzsprung-Russell diagram showing stellar evolution models with initial masses $M_1 = 3.8~M_\odot$ (grey) and $M_1 = 2.7~M_\odot$ (black) along with regions consistent with the radius of the Cepheid, $R_{\rm{Cepheid}} = 17.4\pm 0.9~R_ \odot$ (blue) and that of the red giant companion, $R_{\rm{RG}} = 12.1\pm 2.3~R_\odot$ (red).}\label{f1}
\end{center}
\end{figure}

Given the masses and radii of the two stars in the binary system, we computed stellar evolution models with initial masses, $M_1 = 3.8$ and $M_2 = 2.7~M_\odot$.  The stellar evolution tracks are plotted in Fig.~\ref{f1} along with the regions of the Hertzsprung-Russell diagram consistent with the measured radii. Stellar evolution models appear to fit the measured stellar masses and radii.  However, the models suggest that the Cepheid has an age of about 175~Myr while the red giant star's age is between 420 and 450~Myr. \cite{Piet2011} noted that the stellar ages are approximately 190~Myr and 369~Myr for the Cepheid and red giant, respectively, based on stellar evolution tracks from \cite{Pietrinferni2004}.  These model age differences are due to different amounts of convective core overshooting assumed in the models.  Including overshooting in stellar evolution models lengthens the main sequence lifetime and the greater the amount of overshooting the longer the main sequence lifetime. Regardless of which models are considered, there is a significant age difference. \cite{Piet2011} suggested the binary system may have formed by stellar capture.  We suggest an alternative hypothesis: the apparently younger Cepheid is a stellar merger product between two main sequence stars in what was originally a triple system that has since evolved across the Hertzsprung gap.

 The evolution of the Cepheid in the binary system is also notable because the star appears to be crossing the instability strip for the first time.  This phase of evolution is short, about $10^5$ years for stars with mass of about 3 -- 4~$M_\odot$.  As such, the number of first-crossing Cepheids relative to the number of Cepheids evolving on the blue loop is small, typically about a few percent  \citep{Neilson2012b}. This fact makes CEP1812 more special as a target for understanding stellar evolution and the transition from the main sequence to the red giant branch.

\section{Merger models}
Stellar mergers appear across the HR diagram, from the formation of blue straggler and sdB/O stars \citep{Mateo1990, Schneider2014}, many massive stars \citep{demink2013}, cool R and J stars \citep{Izzard2007, Huang2013}, R~Coronae~Borealis stars \citep{Clayton2013} and anomalous Cepheids \citep{Bono1996}.  The  coalescence of binary companions can occur via orbital disruptions by third bodies \citep{Kozai, Perets2012} or through tidal interactions and Roche Lobe overflow \citep[e.g.][]{Hut1981}.  Stellar mergers occurring  during main sequence evolution quickly settle and evolve in the same manner as a main sequence star formed with mass similar to the sum of the merger progenitors \citep{Glebbeek2013}. This merger rejuvenates the star, making it appear younger, which is consistent with the observed age discrepancy between the Cepheid and its red giant companion in CEP1812.

We propose that the Cepheid component of the eclipsing binary system CEP1812 is the result of a merger between two smaller-mass main sequence stars that have evolved to become a Cepheid.  This progenitor system would have been a hierarchical triple system in which the red giant  was originally the most massive star.  Because of its eccentric orbit (currently about 0.13), the outer star (now the red giant) would have induced Kozai oscillations decreasing the orbital separation between the two smaller-mass stars until tides and Roche lobe overflow dominate the evolution.  The two stars then coalesced rapidly to form a $3.8~M_\odot$ star that appears significantly younger than its companion.

This hypothesis is tested  by computing a stellar evolution model with mass $M_1$ and adding mass $(3.8~M_\odot - M_1)$  after about 310 -- 330 million years of  evolution.   At this age, both stars are evolving on the main sequence. The mass is accreted over a short timescale (relative to the main sequence lifetime of about 10~Myr).  The main effect of accreting mass is to increase the convective core mass of the star by mixing additional hydrogen, hence acts to rejuvenate the star. To first order, this mixing is independent of the merging time scale and given the masses we consider, additional mixing effects are not as important \citep{Glebbeek2013}. The merged star then evolves to the end of red giant evolution beyond the first-crossing of the Cepheid instability strip where Cepheid is observed to be. The initial model is assumed to have mass greater than the accreted companion, i.e. $M_1 > 3.8~M_\odot - M_1$, and  the donor and accretor have the same surface chemical composition.  This is a simplistic calculation that ignores mass lost from the system, which is $\approx 0.1~M_\odot$ \citep{Glebbeek2008}, and ignores issues related to mixing caused by accretion. The test is sufficient, however, to understand whether a stellar merger resolves the age discrepancy.

\begin{figure*}[t]
\begin{center}
\includegraphics[width=0.5\textwidth]{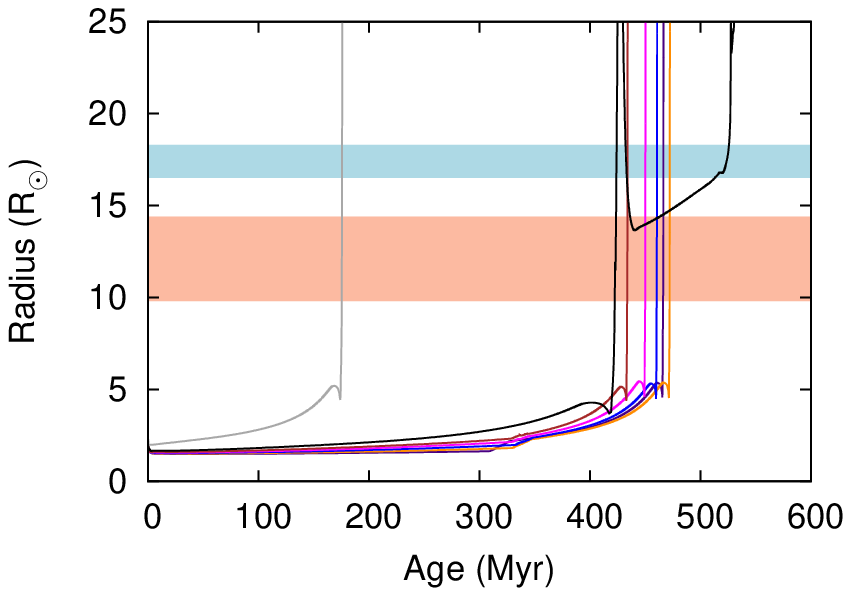}\includegraphics[width=0.5\textwidth]{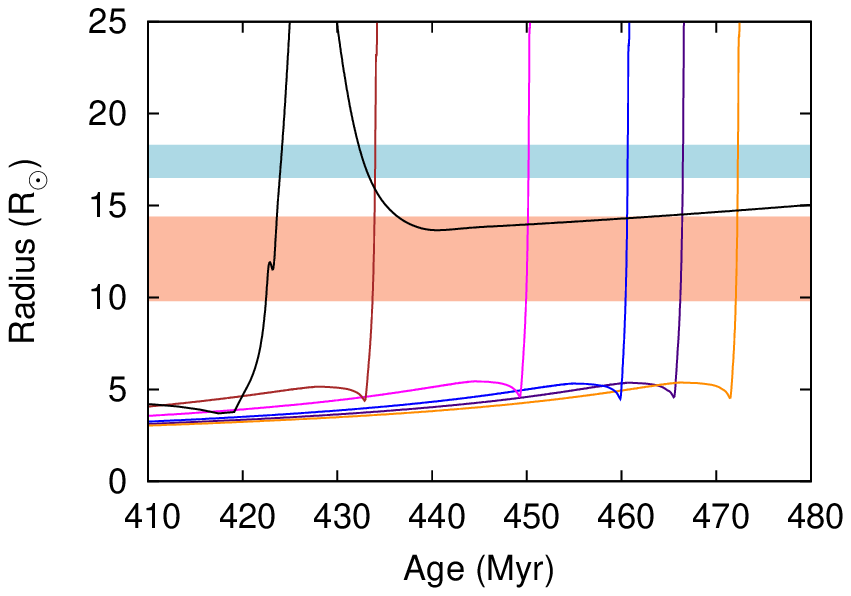}
\caption{(Left) Stellar radii as a function of age for stellar evolution models with initial mass $M_1 = 3.8$ (grey) and $M_2 = 2.7~M_\odot$ (black), along with stellar merger models with progenitor initial masses $2.1$ (violet), $2.2$ (orange), $2.3$ (blue), $2.4$ (pink) and $2.5~M_\odot$ (brown).  The blue and red colored regions represent the measured Cepheid and red giant companion radii, respectively. (Right) A closer view of the stellar radius as a function age consistent with the current age of the red giant}\label{fage}
\end{center}
\end{figure*}

We plot  the stellar radius as a function of age in Fig.~\ref{fage}.  This plot demonstrates the age discrepancy between the $3.8~M_\odot$ Cepheid and its $2.7~M_\odot$ red giant companion.  We also plot stellar merger models with progenitor initial masses $2.1$, $2.2$, $2.3$, $2.4$, and $2.5~M_\odot$ that each accrete mass until they reach $3.8~M_\odot$ and then evolve normally.  Those stellar evolution models are rejuvenated and do not cross the Hertzsprung gap until they are much older than the main sequence life time of stars with that initial mass.  The merger models all have a radius consistent with that of the Cepheid at approximately the same age as the red giant star, implying that CEP1812 is the result of a stellar merger that occurred early in the main sequence evolution of a hierarchical triple system.  However, this solution is not unique, other stellar mass combinations are possible depending on when they merge.

\section{Discussion}
There exists a specific class of Cepheids that appear to have evolved from main sequence stellar mergers: anomalous Cepheids \citep{Bono1996}. They are typically found in older stellar populations in close proximity to RR~Lyrae and horizontal branch stars, primarily in dwarf spheroidal satellite galaxies \citep[e.g.][]{Mateo1995, Kinemuchi2008}.  These stars pulsate with periods ranging from about 0.3 to about 2.5~days and LMC anomalous Cepheids have a predicted average mass $M_{\rm{AC}} = 1.2\pm 0.2~M_\odot$ \citep{Fiorentino2012}. Their masses tend to be smaller than that of LMC Cepheids and there is a small overlap for the longest-period anomalous Cepheid and shortest-period classical Cepheids.  Because of these properties, \cite{Sills2009} suggested they are the merger product between two low-mass main sequence stars that have evolved across the Cepheid instability strip, akin to the models computed in this work, however \cite{Bono1996} showed that a many ACs are low-mass core helium-burning stars  evolving from the horizontal branch. The anomalous Cepheids do not appear to be related to classical Cepheids, but could CEP1812 be a missing link between these two classes of Cepheids?

\begin{figure}[t]
\begin{center}
\includegraphics[width=0.5\textwidth]{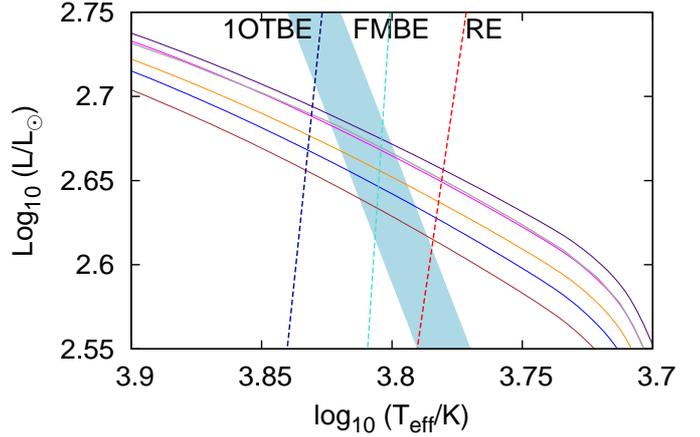}
\caption{Hertzsprung-Russell diagram showing stellar evolution models evolving across the Hertzsprung gap and the first-crossing of the Cepheid instability strip.  Solid lines follow evolution models with initial mass $M_1 = 3.8~M_\odot$ (grey), along with stellar merger models with progenitor initial masses $2.1$ (violet), $2.2$ (orange), $2.3$ (blue), $2.4$ (pink) and $2.5~M_\odot$ (brown).  The blue region represents the measured Cepheid radius while the dashed lines denote the blue edge of the first-overtone instability strip `1OTBE' (dark blue), the boundary between first-overtone and fundamental-mode instability strips `FMBE' (light blue) the the red edge of the Cepheid instability strip `RE' (red) \citep{Bono2000}.}\label{hrd}
\end{center}
\end{figure}

We explore this by plotting our stellar evolution models in a Hertzsprung-Russell diagram in Fig.~\ref{hrd}.  Our merger models evolve across the HR diagram with luminosities consistent with the $3.8~M_\odot$ single star model. This implies that CEP1812 is not likely to be a traditional classical Cepheid, but is  an anomalous Cepheid \citep{Fiorentino2006}.   This possibility is contentious because ACs typically have masses about one solar mass and have lower metallicities.  Because of its mass, CEP1812 likely has a metallicity similar to the assumed LMC metallicity.  The star is also not core helium-burning.  Both properties are inconsistent with various pulsation models of ACs \citep{Bono1996, Caputo2004, Marconi2004}. However,  a massive AC evolving across the Hertzsprung gap would be rare because this stage of evolution is short, less than 1~Myr. 

\cite{Soszynski2008} classified CEP1812 as a classical Cepheid based on Fourier decomposition of the observed light curves \citep{Simon1981}.  Measured Fourier parameters allow for the classification of variable stars because for different variable stars and different pulsation modes the predicted Fourier parameters tend to cluster.  However, Fourier parameters for anomalous Cepheids (ACs) vary significantly and the  anomalous Cepheid Fourier parameters from the OGLE-III survey of the LMC overlap with those of short-period classical Cepheids including CEP1812  \citep{Fiorentino2012}.  The Fourier components for CEP1812 are consistent with both being a short-period classical Cepheid or a first-overtone AC. This is somewhat surprising as the Fourier Component $\phi_{21}$ has only a small range of values for LMC~ACs as measured by \cite{Soszynski2008b}.  This result adds credence to the possibility that CEP1812 is an AC.
\cite{Piet2011} noted that the pulsation period and brightness of CEP1812 is consistent with the measured period-luminosity relation \citep{Soszynski2008}.  However, it is also consistent with the first-overtone AC period-luminosity relation  \citep{Ripepi2014}.   Both its pulsation period and amplitude are consistent with that of other LMC~ACs in the OGLE-III survey \citep{Soszynski2008b} again suggesting that CEP1812 may be an AC and not a classical Cepheid.
Based on our models and the fact that CEP1812 is consistent with the pulsation properties of other LMC anomalous Cepheids, we suggest that CEP1812 is not a classical Cepheid but is instead an anomalous Cepheid.  Thus, CEP1812 is the most massive AC discovered and is a factor of two more massive than those in  the \cite{Fiorentino2012} sample, where the mass is measured from period-luminosity-color relations.  

It is not surprising that massive anomalous Cepheids are rare.  Stellar mergers are, themselves, rare, but the most noticeable difference between a single star evolving and a stellar merger is that the latter may appear to be younger.  Examples are blue straggler stars which are easy to detect in a Globular cluster because the majority of stars there are very old.  ACs are detected by the same method: they tend to be found in old populations along with horizontal branch and RR~Lyrae stars. It is this contrast that allows anomalous Cepheids to be detected and that also explains why they all tend to be about one solar mass.  Only low-mass stars have life times long enough to both undergo a stellar merger and appear in old populations, massive stars that merge would have disappeared long before the general population had significantly evolved. A merger between a 2.4 and a $1.4~M_\odot$ star would not appear significantly rejuvenated relative to nearby field stars,  but does stand out in a binary system.

Although we suggest that CEP1812 could be classified as an AC, it is not clear that the star will have any properties significantly different from other classical Cepheids.  \cite{Langer2012} suggested that strong magnetic fields, typically about 1~kG, that are found in about 10\% of intermediate-mass and massive main sequence stars \citep{Donati2009} could be generated in a main sequence merger.  If this hypothesis is correct then CEP1812 could presently have a weak magnetic field that might affect various pulsation properties.   Another test of our merger scenario is the abundance of CEP1812. If the star is the product of a merger of two main sequence stars, where the mass donor has underwent some hydrogen burning then one might expect some chemical anomalies, particularly in the surface helium, nitrogen and carbon abundances. However, because this is a low-mass merger, any changes to the carbon and nitrogen abundances will be small, about a fraction of a dex \citep{Glebbeek2013}.

Because the star is a merger product,  one might expect some abundance anomalies.  However, our model assumes that the merger is a product of two stars with the same initial composition, hence any anomalies would be small.  While CEP1812 is a merger of two main sequence stars, some anomalous Cepheids might be created by the merger between a main sequence star and an evolved helium-burning star which would generate a new helium-burning star with an envelope that is massive relative to the total stellar mass, or conversely the core mass is much smaller than for a single star at a similar stage of evolution \citep{demink2013}.  These post-main sequence mergers could also evolve to become Cepheids but will have very different properties because they have a smaller luminosity for a given mass and a more massive envelop leading to different pulsation properties.

In summary, CEP1812 may be an anomalous Cepheid that appears to be just like any classical Cepheid such that it may be the missing link between the two classes of stars.  However, the strange evolution of CEP1812 implies that it might be unwise to use this star as a calibrator for the Cepheid Leavitt Law or for resolving the Cepheid mass discrepancy.

\acknowledgements
This work has been supported by a research grant from the NSF (AST-0807664). RGI acknowledges funding from the Alexander von Humboldt Foundation and the Science \& Technology Facilities Council.

\bibliographystyle{aa} 

\bibliography{c1812} 

\end{document}